\documentclass[aps,prl,superscriptaddress,twocolumn,floatfix]{revtex4-1}
\usepackage{units}
\usepackage{amsmath}
\usepackage{amssymb}
\usepackage{graphicx}
\usepackage{bm}
\usepackage{multirow,color,relsize,ulem,microtype,bbold}

\newcommand{\be}{\begin{equation}}
\newcommand{\ee}{\end{equation}}

\newcommand{\re}[1]{\text{Re}[#1]}

\newcommand{\ep}{{\scriptscriptstyle \text{EP}}}

\begin{document}
\title{Anomalous Parity-Time Symmetry Transition away from an Exceptional Point}

\author{Li Ge}
\email{li.ge@csi.cuny.edu}
\affiliation{\textls[-18]{Department of Engineering Science and Physics, College of Staten Island, CUNY, Staten Island, NY 10314, USA}}
\affiliation{The Graduate Center, CUNY, New York, NY 10016, USA}

\date{\today}

\begin{abstract}
Parity-time ($\cal PT$) symmetric systems have two distinguished phases, e.g., one with real energy eigenvalues and the other with complex conjugate eigenvalues. To enter one phase from the other, it is believed that the system must pass through an exceptional point, which is a non-Hermitian degenerate point with coalesced eigenvalues and eigenvectors. In this letter we reveal an anomalous $\cal PT$ transition that takes place away from an exceptional point in a nonlinear system: as the nonlinearity increases, the original linear system evolves along two distinct $\cal PT$-symmetric trajectories, each of which can have an exceptional point. However, the two trajectories collide and vanish away from these exceptional points, after which the system is left with a $\cal PT$-broken phase. We first illustrate this phenomenon using a coupled mode theory and then exemplify it using paraxial wave propagation in a transverse periodic potential.
\end{abstract}

\maketitle

Parity-time ($\cal{PT}$) symmetry originated in the search for an alternative framework of canonical quantum mechanics and quantum field theory \cite{Bender1,Bender2,Bender3}. It has since stimulated fast growing interest in optics \cite{El-Ganainy_OL07,Moiseyev,Musslimani_prl08,Makris_prl08,Guo,mostafazadeh,Longhi,CPALaser,conservation,Ge_PRX2014,Ruter,Lin,Feng,Feng2,Walk,Hodaei,Yang},
microwaves \cite{Microwave}, radio waves \cite{RC}, acoustics \cite{sound}, and mechanics \cite{mechanics}. In all these systems, a well-known and intriguing property is the existence of two distinguished phases, e.g., one with real energy eigenvalues (``$\cal PT$-symmetric phase") and the other with complex conjugate eigenvalues (``$\cal PT$-broken phase"). The same property is shared with other systems with novel symmetries \cite{antiPT}, which is the consequence of having a pseudo-Hermitian Hamiltonian \cite{PseudoH}.

The two aforementioned phases are separated by exceptional points (EPs) \cite{EP1,EP2,EP3,EPMVB,EP4,EP5,EP6,EP7,EP8,EP9,EP10}, which are non-Hermitian degenerate points with coalesced eigenvalues and eigenvectors.
While EPs are ubiquitous in non-Hermitian systems, they are singular points in the parameter space and can be reached only by a sweep involving two or more parameters in general. $\cal PT$-symmetric systems are special in this regard, as they only require sweeping a single parameter to reach an EP. This parameter can be, for example, the gain and loss strength in the system or the effective wavelength of the eigenstates \cite{multimode}. As such, it is believed that if the system maintains $\cal PT$ symmetry, then it must pass through an EP in order to enter one phase from the other, regardless of which parameter is varied. To the best of our knowledge, the only exception to this rule occurs when the underlying Hermitian system (i.e., without gain or loss) has genuine degeneracy \cite{Ge_PRX2014,Feng2} with identical eigenvalues but distinct eigenstates. This scenario, nevertheless, can be taken as the limiting case of a system with an EP and increasing system size \cite{barRings}.

In this letter, we reveal an anomalous transition from the $\cal PT$-symmetric phase to the $\cal PT$-broken phase that takes place \textit{away} from an EP in a nonlinear system:  as the nonlinearity increases, the original linear system evolves along two distinct $\cal PT$-symmetric trajectories, each of which can have an EP. However, the two trajectories collide and vanish away from these EPs, after which the system is left with only a $\cal PT$-broken phase.

We will refer to this phenomenon as anomalous $\cal PT$ transition (APT).
Below we first illustrate the existence of APT using a coupled mode theory, and we show that APT cannot be induced with the typical form of nonlinearity considered previously, i.e., with identical energy shift coefficients in the coupled systems \cite{Ramezani,Graefe}. Instead, APT requires distinct and eigenstate-dependent paths of the effective Hamiltonian as the nonlinearity increases, which we illustrate using nonlinearity-shifted couplings. We then exemplify APT using paraxial wave propagation in a transverse periodic potential, and in the conclusion we discuss how APT can be identified in an experiment and show that it does not occur from the $\cal PT$-broken phase to the $\cal PT$-symmetric phase.

We start our discussion by considering two identical oscillators with energy $E_0$ and a real-valued coupling $g_0$ (which can be negative). One oscillator is subjected to gain at rate $+i\kappa_0$, and the other is subjected to loss at rate $-i\kappa_0$. Before we introduce nonlinearity, the effective Hamiltonian of the system can be written as
\be
H_0 = \begin{bmatrix}
E_0 + i\kappa_0 & g_0 \\
g_0 & E_0 - i\kappa_0
\end{bmatrix},\label{eq:CMT0}
\ee
which is $\cal PT$-symmetric and well studied. It satisfies ${\cal PT}H_0{\cal PT}=H_0$, where the parity operator $\cal P$ is represented by a rotation matrix $[\begin{smallmatrix} 0 & 1 \\ 1& 0 \end{smallmatrix}]$ and the time-reversal operator $\cal T$ by the complex conjugate.
The two eigenvalues of $H_0$ is given by $E^{(1,2)}=E_0\pm\sqrt{g_0^2-\kappa_0^2}$, which are real when $|g_0|>\kappa_0$ and the system is in the $\cal PT$-symmetric phase; they form a complex conjugate pair when $|g_0|<\kappa_0$ and the system is in the $\cal PT$-broken phase. The EP is located at $|g_0|=\kappa_0$, which the system must pass through to go from one phase to the other.

The two eigenstates of the system can be expressed as $\psi^{(j)}=c^{(j)}_a\varphi_a + c_b^{(j)}\varphi_b\,(j=1,2)$, where $\varphi_{a,b}$ are the uncoupled wave functions of the two oscillators. Below we drop the superscript $j$ when ambiguity is unlikely, and we use the normalization $|c_a|^2+|c_b|^2\equiv1$ as usual. We also note that $|c_a|$ and $|c_b|$ are equal in the $\cal PT$-symmetric phase (given by $1/\sqrt{2}$), which is not the case in the $\cal PT$-broken phase [see the Supplemental Information (SI)]. These properties play an important role in our analysis below.

To illustrate the simplest case where APT arises, we take the gain and loss strength $\kappa_0$ to be independent of the nonlinearity $\varepsilon$. %The latter is positive (negative) for defocusing (focusing) nonlinearity
We assume the typical nonlinear energy shift in the effective Hamiltonian, with $E_0$ replaced by $E_{a,b}(\varepsilon)=E_0+2\varepsilon|c_{a,b}|^2$ in the two diagonal elements \cite{Ramezani,Graefe}. Most importantly, we consider nonlinearity-shifted couplings given by
\begin{align}
g_{a}(\varepsilon)&=g_0 + \varepsilon\beta c_{a}^*c_{b} + \varepsilon\gamma|c_a|^2, \label{eq:ga} \\
g_{b}(\varepsilon)&=g_0 + \varepsilon\beta c_{b}^*c_{a} + \varepsilon\gamma|c_b|^2,\label{eq:gb}
%\kappa_{a,b}(\varepsilon)=\kappa_0 + \varepsilon \beta_\kappa c_{a,b}^*c_{b,a} + \varepsilon \gamma_\kappa |c_{a,b}|^2.
\end{align}
which are the key quantities for APT to take place as we will show. Here $\beta,\gamma$ are two real constants, and $g_a=g_b^*$ holds by construction when $|c_a|=|c_b|$. The global phase of $\psi$, which does not bear a physical significance, is eliminated in $g_{a,b}(\varepsilon)$ thanks to the products $c_a^*c_b$ and $c_b^*c_a$.
Below we will refer to our nonlinear Hamiltonian as
\be
H \equiv \begin{bmatrix}
E_a(\varepsilon) + i\kappa_0 & g_a(\varepsilon) \\
g_b(\varepsilon) & E_b(\varepsilon) - i\kappa_0
\end{bmatrix},\label{eq:CMT_nonlinear}
\ee
and we recover the typical nonlinear Hamiltonian mentioned previously ($\tilde{H}$) when $\beta,\gamma$ are taken as zero (i.e., $g_{a,b}(\varepsilon)=g_0$).

While $\tilde{H}$ displays interesting dynamical effects \cite{Ramezani}, it does not lead to a transition from the $\cal PT$-symmetric phase to the $\cal PT$-broken phase when the nonlinearity strength $|\varepsilon|$ increases from zero. This observation can be seen in the following way. Let us start in the $\cal PT$-symmetric phase with $|g_0|>\kappa_0$. As mentioned previously, $|c_a|=|c_b|=1/\sqrt{2}$ holds for both linear eigenstates. As $|\varepsilon|$ increases, $c_{a,b}$ evolve continuously from their linear values, and if we assume that they still have the same modulus, then $\tilde{H}$ is simply $H_0+\varepsilon \bm{1}$, where $\bm{1}$ is the identity matrix. Therefore, the two eigenvalues shift in parallel, i.e., $E^{(1,2)}(\varepsilon)=E^{(1,2)}(0)+\varepsilon$, and they will not be able to coalesce and enter the $\cal PT$-broken phase when $\varepsilon$ varies. In the meanwhile, the eigenstates are unchanged, which is consistent with our assumption that $|c_a|=|c_b|$. We note that
these two nonlinear eigenstates are the only ones originating from the linear eigenstates, even though additional nonlinear eigenstates can appear elsewhere \cite{Graefe}. In conclusion, the $\cal PT$-symmetric phase persists despite the increasing nonlinearity.

\begin{figure}[t]
\includegraphics[clip,width=\linewidth]{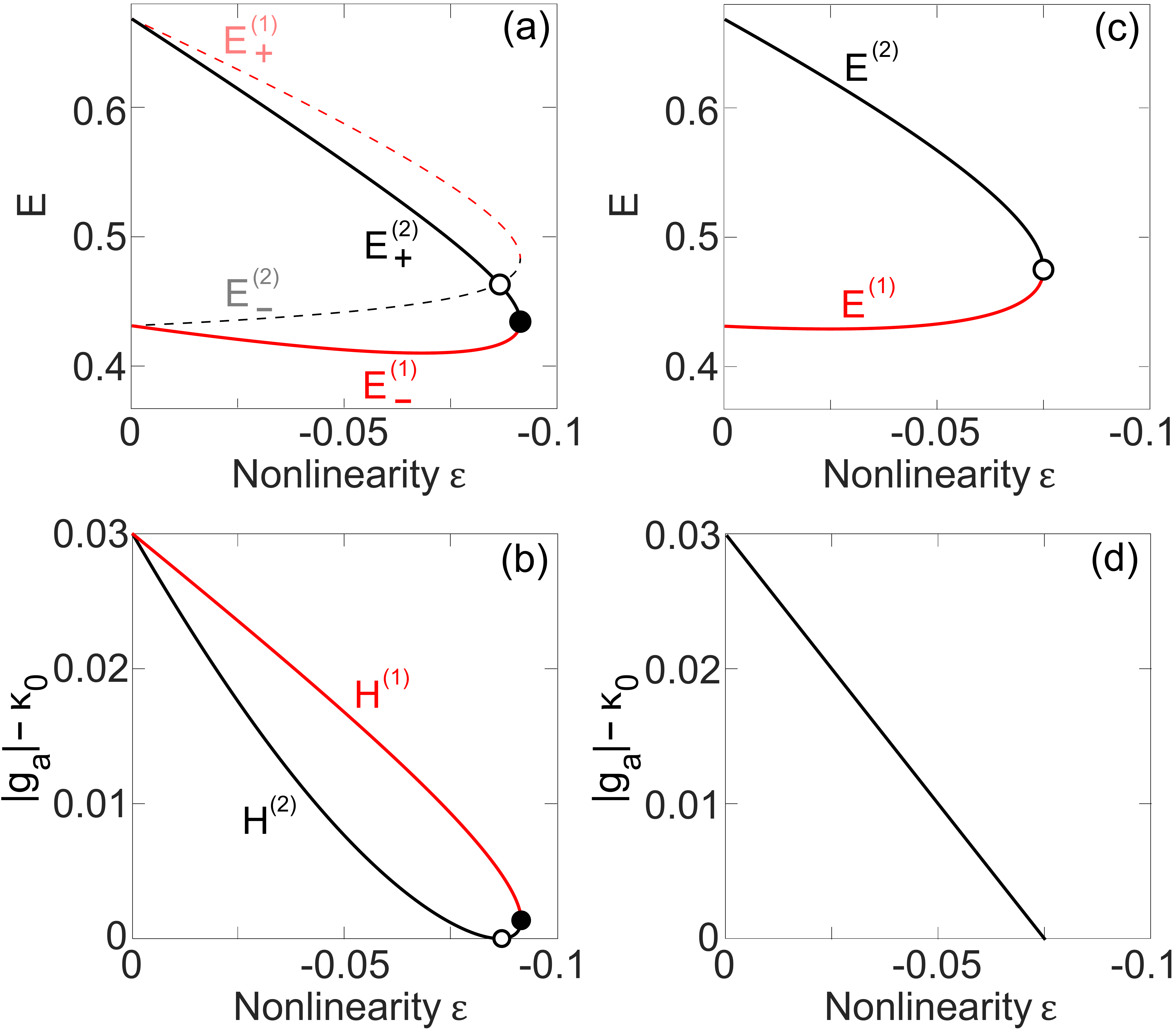}
\caption{(Color online) Anomalous $\cal PT$ transition away from an EP. (a) Two nonlinear energy eigenvalues $E^{(1)}_-$ and $E^{(2)}_+$ in the $\cal PT$-symmetric phase (solid lines) annihilate each other at the APT point where $\varepsilon=-0.091$ (filled circle). Dashed lines show the two additional eigenvalues $E^{(1)}_+$ and $E^{(2)}_-$ of the linearized Hamiltonians $H^{(1,2)}$, and the open circle shows the EP of $H^{(2)}$. (b) Difference between $|g^{(j)}_a(\varepsilon)|$ and $\kappa_0$. Filled and open circles show the APT point and EP, respectively. The parameters used in (a) and (b) are: $E_0 = 0.55$, $\kappa_0=0.22$, $g_0=-0.25$, $\beta=0.6$, and $\gamma=-0.8$. (c) Same as (a) but with $\beta=0$. The APT point is replaced by an EP at $\varepsilon=-2(\kappa_0+g_0)/\gamma=-0.075$. The difference between the now path-independent $|g_a(\varepsilon)|$ and $\kappa_0$ is shown in (d).}\label{fig:nonlinear_cmt}
\end{figure}

In contrast, the nonlinear Hamiltonian $H$ given by Eq.~(\ref{eq:CMT_nonlinear}) displays a qualitatively different behavior. Similar to our discussion of $\tilde{H}$, we start in the $\cal PT$-symmetric phase and again assume that $|c_a|=|c_b|$ holds after nonlinearity is introduced to the system. Now besides $E_a(\varepsilon)=E_b(\varepsilon)$, we also find $g_a^{(j)}(\varepsilon)=[g_b^{(j)}(\varepsilon)]^*$ as mentioned previously. It is important to note that the couplings $g^{(1,2)}_{a}(\varepsilon)$ [and $g^{(1,2)}_{b}(\varepsilon)$] differ, which prompts us to restore the nonlinear mode index $j$ ($j=1,2$). This $j$-dependence arises from the product $c_a^*c_b$ in Eq.~(\ref{eq:ga}), or equivalently the relative phase between $c_a$ and $c_b$, which is different for the two eigenstates (see SI). This $j$-dependence, or equivalently a nonzero $\beta$,  leads to APT as we shall see.

Along these two $j$-dependent nonlinear trajectories, the system now has two distinct linearized Hamiltonians $H^{(j)}$, each of which is still $\cal PT$-symmetric and satisfies ${\cal PT}H^{(j)}{\cal PT}=H^{(j)}$. The eigenvalues of $H^{(j)}$ are hence either real or complex conjugates, and they are given by
\be
E^{(j)}_\pm=E_0 + \varepsilon \pm \sqrt{|g^{(j)}_a(\varepsilon)|^2-\kappa_0^2}. \label{eq:E_CMT_nonlinear}
\ee
The corresponding eigenvectors in the $\cal PT$-symmetric phase still satisfy $|c_a|=|c_b|$, which is again consistent with our assumption. We note that the two linearized Hamiltonians $H^{(1,2)}$ have four eigenvalues in total, but for each $H^{(j)}$, only one of its two eigenvalues given by Eq.~(\ref{eq:E_CMT_nonlinear}) corresponds to the nonlinear eigenstate $\psi^{(j)}$. These nonlinear eigenstates are stable (see SI), and we denote their eigenvalues by $E^{(1)}_-,E^{(2)}_+$, with the other two spurious ones by $E^{(1)}_+,E^{(2)}_-$ [see Fig.~\ref{fig:nonlinear_cmt}(a)].

\begin{figure}[t]
\includegraphics[clip,width=\linewidth]{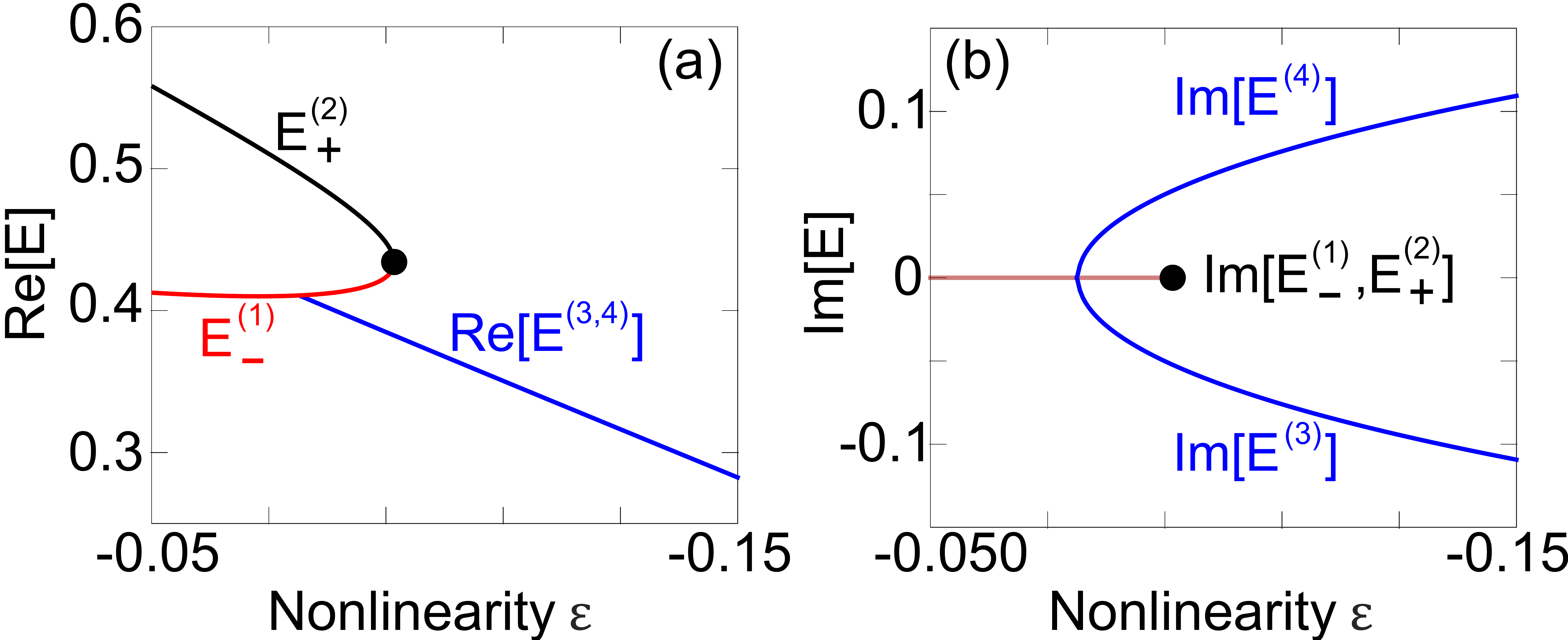}
\caption{(Color online) Additional nonlinear eigenvalues $E^{(3,4)}$ not shown in Fig.~\ref{fig:nonlinear_cmt}(a). They make up the $\cal PT$-broken phase beyond the APT point (dots).}\label{fig:nonlinear_cmt_b}
\end{figure}

As is clear from Eq.~(\ref{eq:E_CMT_nonlinear}), each of the two $H^{(j)}$ can have an EP at $|g^{(j)}_a(\varepsilon)|=\kappa_0$, which could in principle lead to two $\cal PT$ transitions to their respective $\cal PT$-broken phases. However, APT takes place away from these EPs, when $E^{(1)}_-$ and $E^{(2)}_+$ annihilate each other at a different nonlinearity strength [see Fig.~\ref{fig:nonlinear_cmt}(a)]. We will refer to this annihilation point as the APT point, beyond which the system is left with only a $\cal PT$-broken phase, which we will discuss later in Fig.~\ref{fig:nonlinear_cmt_b}.

We have labeled $E^{(2)}_{\pm}$ by continuity beyond their EP in Fig.~\ref{fig:nonlinear_cmt}(a), i.e., with inverted signs before the square root in Eq.~(\ref{eq:E_CMT_nonlinear}). It is straightforward to see from Eq.~(\ref{eq:E_CMT_nonlinear}) that the annihilation of $E^{(1)}_-$ and $E^{(2)}_+$ is accompanied by $|g_a^{(1)}(\varepsilon)|=|g_a^{(2)}(\varepsilon)|$. In fact not just their moduli, $g_a^{(1,2)}(\varepsilon)$ themselves also become the same at the APT point. As we show in SI, they are given by the intersections of a circle and a hyperbola in the complex plane, both parametrized by $\varepsilon$.
These two conic curves become tangent to each other at a maximum nonlinearity strength $|\varepsilon|_\text{max}$, beyond which they no longer intersect.
$|\varepsilon|_\text{max}$ gives the horizontal position of the APT point, and it is $0.091$ in the example shown in Figs.~\ref{fig:nonlinear_cmt}(a) and \ref{fig:nonlinear_cmt}(b).

To verify that the APT point is not an EP itself, we compute the difference between $|g^{(j)}_a(\varepsilon)|$ and $\kappa_0$ along the two nonlinear trajectories. As Fig.~\ref{fig:nonlinear_cmt}(b) shows, this difference diminishes as $\varepsilon$ reduces, but it does not become zero at the APT point, where %$g_a^{(1,2)}=-0.220-0.027i$ and
$|g_a^{(1,2)}|-\kappa_0=1.36\times10^{-3}$. In fact, this difference reaches zero at an EP along the trajectory of $H^{(2)}$ before the APT point. It may look surprising at first as to why $E^{(2)}_\pm$ come right back into the $\cal PT$-symmetric phase beyond this EP instead of entering the $\cal PT$-broken phase. However, one quickly realizes that since $E^{(1)}_-$ is still in the $\cal PT$-symmetric phase beyond this EP, $E^{(2)}_+$ has to stay in the $\cal PT$-symmetric phase also in order to annihilate it at the APT point, where they are both real. In this sense, it is the APT that prevents $E^{(2)}_\pm$ from entering the $\cal PT$-broken phase beyond its EP. In addition,  we note  that $\kappa_0$ is not just the cut-off of $|g^{(2)}_a(\varepsilon)|$ imposed by the $\cal PT$-symmetric phase; it is also the true minimum of $|g^{(2)}_a(\varepsilon)|$ which cannot be passed. This is evidenced by the vanishing slope of $|g^{(2)}_a(\varepsilon)|$ at the APT point shown in Fig.~\ref{fig:nonlinear_cmt}(b), and we provide a proof in SI.
We also note that the EP before the APT point can occur on the trajectory of $H^{(1)}$ instead, if $|g^{(1)}_a(\varepsilon)|<|g^{(2)}_a(\varepsilon)|$ in the $\cal PT$-symmetric phase (see SI).

%The next question is whether this EP always appears before the APT point. The answer is affirmative, and to prove this claim we first establish the following facts. First, one $|g^{(j)}_a|$ is always greater than the other (i.e., $g^{(3-j)}_a$) before the APT point [see Fig.~\ref{fig:nonlinear_cmt}(b)], at which they become the same. Here $j=2$ but the opposite can also happen (see SI). Second, according to Eq.~(\ref{eq:E_CMT_nonlinear}) $E^{(j)}_+$ ($E^{(j)}_+$) is then greater (smaller) than both $E^{(3-j)}_\pm$ before the ...

The annihilation of two eigenvalues is a generic feature in non-Hermitian systems and in nonlinear systems upon the variation of a parameter. Here this tuning parameter is the nonlinearity itself, and other instances can be, for example, the lengths of the gain and loss regions in a slab laser \cite{EP7} and a random laser \cite{Andreasen}. In fact, this annihilation also happens when $\beta=0$ [see Fig.~\ref{fig:nonlinear_cmt}(c)], with which $g_{a,b}(\varepsilon)$ no longer depend on the nonlinear mode index $j$ in the $\cal PT$-symmetric phase: they only depend on $|c_a|$ and $|c_b|$, which are the same (i.e., $1/\sqrt{2}$) for the two nonlinear eigenstates. As a result, these two nonlinear states $\psi^{(1,2)}$ are captured by the same linearized Hamiltonian $H$, and their eigenvalues are given by $E^{(1,2)}=E_0 + \varepsilon \pm \sqrt{|g_a(\varepsilon)|^2-\kappa_0^2}$. Therefore, if these two nonlinear eigenstates annihilate, it has to be at an EP where $|g_a(\varepsilon)|=\kappa_0$ [see Fig.~\ref{fig:nonlinear_cmt}(d)]. From this comparison we see that a nonzero $\beta$, or more generally, a path-dependent evolution of $g_{a,b}$ and $H$, leads to the occurrence of APT.

As to the $\cal PT$-broken phase beyond the APT point, it consists of two additional nonlinear eigenstates $\psi^{(3,4)}$ that spin off from one of the two $\cal PT$-symmetric eigenstates (see Fig.~\ref{fig:nonlinear_cmt_b}). %The system has a coexisting $\cal PT$-symmetric phase (due to $\psi^{(1,2)}$) and a $\cal PT$-broken phase (due to $\psi^{(3,4)}$) for a nonlinearity strength between the emerging point of $\psi^{(3,4)}$ and the APT point.
We note that if $\psi^{(3)}=c_a\varphi_a + c_b\varphi_b$ is a nonlinear eigenstate, it is straightforward to show that $c_b^*\varphi_a + c_a^*\varphi_b$ is also a nonlinear eigenstate of $H$ given by Eq.~(\ref{eq:CMT_nonlinear}). This is indeed how
$\psi^{(3,4)}$ are related, i.e. they satisfy ${\cal PT}\psi^{(4)}=\psi^{(3)}(x)$, and their eigenvalues satisfy $E^{(4)}=[E^{(3)}]^*$. These properties are identical to those in a linear $\cal PT$-broken phase, but we emphasize that $\psi^{(3,4)}$ are eigenstates of two distinct linearized Hamiltonian $H^{(3,4)}$, respectively. Neither of $H^{(3,4)}$ is $\cal PT$-symmetric, i.e., ${\cal PT}H^{(3,4)}{\cal PT}\neq H^{(3,4)}$, but they are $\cal PT$-symmetric partners and satisfy ${\cal PT}H^{(3)}{\cal PT}=H^{(4)}$. $H^{(3,4)}$ each has an additional eigenvalue that does not exist in the nonlinear system (not shown), similar to $H^{(1,2)}$ in the $\cal PT$-symmetric phase.

\begin{figure}[t]
\includegraphics[clip,width=\linewidth]{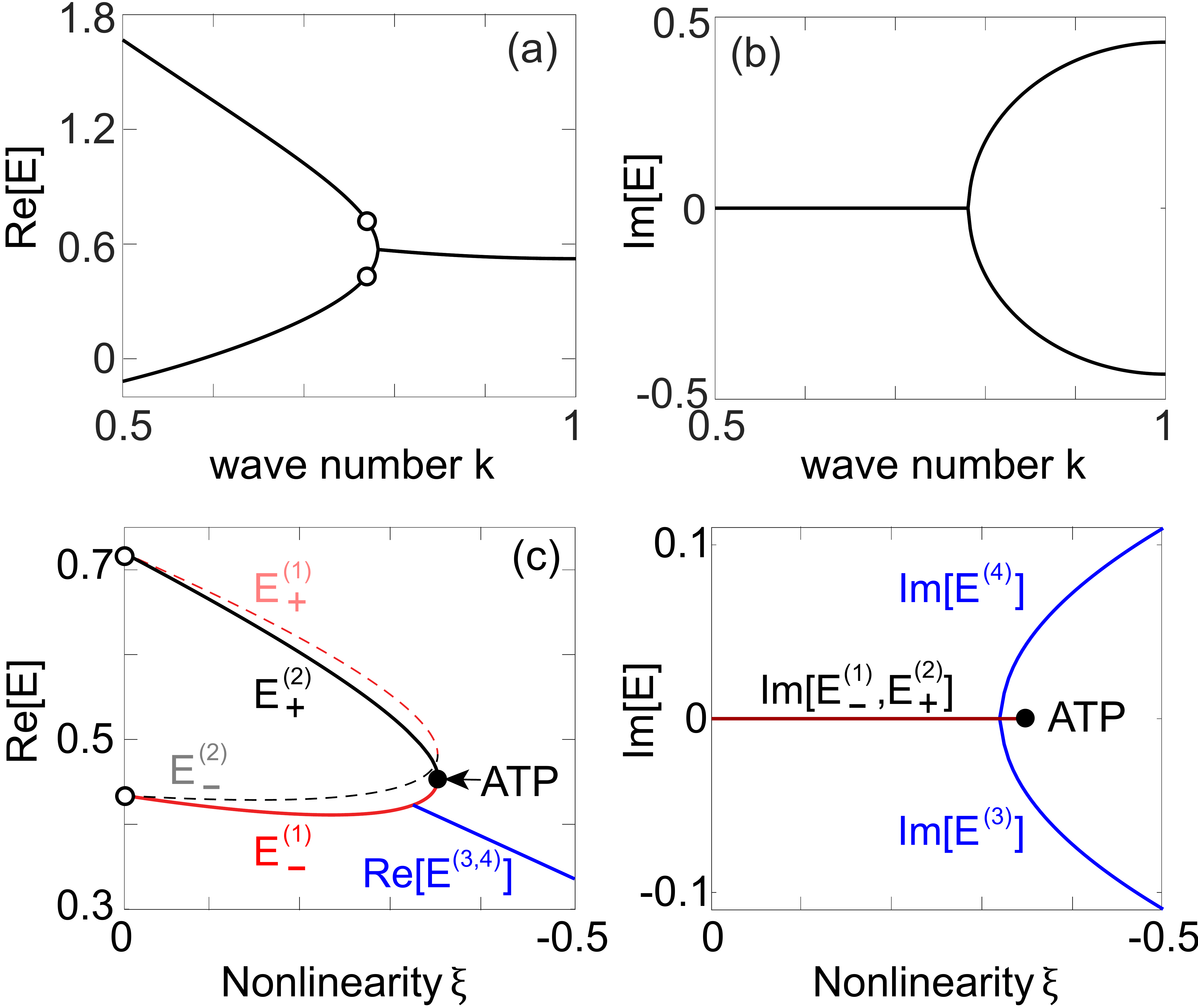}
\caption{(Color online) Anomalous $\cal PT$ transition in a transverse periodic potential $V_0(x)=-\cos(x)^2-i\sin(2x)$ with Kerr nonlinearity. (a,b) Real and imaginary parts of the first two band near the band edge $k=1$ (solid lines) in the linear regime. Open circles in (a) and (c) mark the same pair of modes at $k=0.77$ that we study in the nonlinear regime. (c,d) Anomalous $\cal PT$ transition from the $\cal PT$-symmetric phase to the $\cal PT$-broken phase when nonlinearity increases, similar to that shown in Figs.~\ref{fig:nonlinear_cmt}(a) and (b).}\label{fig:periodic}
\end{figure}

To exemplify APT in a model system, we consider paraxial wave propagation with Kerr nonlinearity
\be
i\partial_z \psi \equiv H \psi = -\partial^2_x \psi + V_0(x)\psi + \xi|\psi|^2\psi,\label{eq:paraxial}
\ee
where $\psi(x,z)$ is the wave function normalized by $\langle\psi|\psi\rangle\equiv\int_{-L/2}^{L/2} |\psi|^2\,dx=1$ and $z,x$ are the scaled coordinates of the longitudinal and transverse directions. $L$ is the length of one period of the potential $V_0(x)=V_R(x)+iV_I(x)$, which is $\cal PT$-symmetric and satisfies $V_R(-x)=V_R(x)$ and $V_I(-x)=-V_I(x)$. For simplicity, we consider $V_R(x)=-\cos(x)^2$ and $V_I(x)=-\tau\sin(2x)$, which have been studied previously in the linear regime \cite{Makris_prl08}.
Its first two linear bands (with $\xi=0$) are in the symmetric phase unless $|\tau|$ is larger than $0.5$ \cite{Makris_prl08}, with which the modes near the band edge enter the $\cal PT$-broken phase [see Figs.~\ref{fig:periodic}(a) and \ref{fig:periodic}(b)].

In Figs.~\ref{fig:periodic}(c) and \ref{fig:periodic}(d) we focus on the two modes $\psi^{(1,2)}(x)$ at $k=0.77$, which are in the linear $\cal PT$-symmetric phase with $\tau=1$. We note that the intensities of these two modes satisfy $|\psi^{(j)}(-x)|^2=|\psi^{(j)}(x)|^2$, which is equivalent to $|c_a|^2=|c_b|^2$ in the coupled mode theory discussed previously. As a result, they do not break the $\cal PT$-symmetry of the system, since now the nonlinearity-modified potential $V^{(j)}(x)=V_0(x)+\xi|\psi^{(j)}(x;k)|^2$ still has a symmetric real part (i.e., $V_R(x)+\xi|\psi^{(j)}(x;k)|^2$) and an antisymmetric imaginary part (i.e., $V_I(x)$). As we have emphasized in the coupled mode theory, APT requires two path-dependent evolutions of the system Hamiltonian with nonlinearity. This property is satisfied here because $|\psi^{(1)}(x)|^2\neq|\psi^{(2)}(x)|^2$ in the linear case, resulting in different nonlinear potentials $V^{(j)}(x)$ and path-dependent $H^{(1,2)}$.

By choosing a focusing nonlinearity ($\xi<0$) and increasing its strength, we find that $\psi^{(1,2)}(x)$ indeed display APT [see the solid lines in Fig.~\ref{fig:periodic}(c) and \ref{fig:periodic}(d)]: they approach each other and annihilate at $\xi = -0.35$, beyond which the system is left with a $\cal PT$-broken phase. Similar to the situation in the coupled-mode theory, each linearized $H^{(j)}$ has more than one eigenstate, but only one of them corresponds the nonlinear mode $\psi^{(j)}$. The others nevertheless indicate where the EP of $H^{(j)}$ is. As can be seen from Fig.~\ref{fig:periodic}(c), the EP of $H^{(2)}$ (where $E^{(2)}_\pm$ crosses) is again located at a smaller nonlinearity strength than the APT point, similar to the scenario shown in Fig.~\ref{fig:nonlinear_cmt}(a).
%In the $\cal PT$-broken phase, if one finds one nonlinear solution $\psi^{(3)}(x)$ and the resulting $H^{(3)}(x)=-\partial_x^2+V_0(x)+\xi|\psi^{(3)}(x)|^2$, we know immediately that another nonlinear solution $\psi^{(4)}(x) = [\psi^{(3)}(-x)]^*$ also exists, which is one eigenstate of $H^{(4)}(x)=-\partial_x^2+V_0(x)+\xi|\psi^{(4)}(x)|^2=[H^{(3)}(-x)]^*$.
In SI we formulate a two-mode coupled mode theory that reproduces the APT in this example. Although the nonlinearity does not take the exact form as Eq.~(\ref{eq:CMT_nonlinear}), we note that $g_b=g_a^*$ still holds in the nonlinear $\cal PT$ symmetric phase, where the resulting Hamiltonian itself is $\cal PT$-symmetric as well as path-dependent. These two conditions are crucial for APT as we have shown, and they can also be realized, for example, with a gain and loss strength that depends on the nonlinear eigenstates.

%\begin{figure}[b]
%\includegraphics[clip,width=\linewidth]{Fig4.pdf}
%\caption{A nonlinear $\cal PT$ transition from the $\cal PT$-broken phase to the $\cal PT$-symmetric phase for the same periodic potential in Fig.~\ref{fig:periodic}. (a) and (b) show $\im{E}$ and $\re{E}$ as a function of $\xi$ at $\tau=1$ and $k=0.785$. It does not follow the APT mechanism. }\label{fig:periodic_nonlinear2}
%\end{figure}

In summary, we have revealed an anomalous $\cal PT$ transition from the $\cal PT$-symmetric phase to the $\cal PT$-broken phase that takes place away from an EP. We note that the transition in the opposite direction is not an APT: for two $\cal PT$-broken eigenstates with complex conjugate eigenvalues to coalesce, they must become real simultaneously at some nonlinearity strength, which is an EP by definition. Hence this transition (see Ref.~\cite{Segev}, for example) follows the standard $\cal PT$ transition mechanism. It may look difficult to distinguish APT from a standard $\cal PT$ transition (cf. Figs.~\ref{fig:nonlinear_cmt}(a) and (c)) in an experiment, because the spurious eigenvalues $E^{(1)}_+,E^{(2)}_-$ of the linearized Hamiltonians cannot be accessed to identify the EP. One possibility to overcome this difficulty is to prepare another ``conjugate" system, where the sign of $\beta$ is flipped. As we show in SI, APT is not affected by flipping the sign of $\beta$, and it still occurs at the same nonlinearity $\varepsilon$. In addition, the nonlinear eigenvalues of this conjugate system are exactly the spurious eigenvalues $E^{(1)}_+,E^{(2)}_-$ in the original system, and the crossing of these two sets of nonlinear eigenvalues gives the EP.

%Finally, although $g_{a,b}$ are introduced in Eqs.~(\ref{eq:ga}) and (\ref{eq:gb}) as nonlinearity-shifted couplings, $\varepsilon$ can also be interpreted as an external parameter in a linear system that can be varied independently (see SI). Therefore, we expect that APT can also take place in a linear system.

We thanks Ramy El-Ganainy for helpful discussions. This project is supported by PSC-CUNY Grant No. 68698-0046 and NSF Grant No. DMR-1506987.

\section*{Supplemental Information}

\subsection{Eigenstates in the coupled-mode theory}

The eigenstates $[c_a\; c_b]^T$ of the effective Hamiltonian $H_0$ given by Eq.~(\ref{eq:CMT0}) in the main text satisfy
\be
c_b = \frac{\pm\sqrt{g_0^2-\kappa_0^2}-i\kappa_0}{g_0}c_a. \label{eq:c1c2}
\ee
Here ``$T$" denotes the matrix transpose. In the $\cal PT$-symmetric phase $|g_0|>\kappa_0$, and the expression above implies $|c_a|^2=|c_b|^2$. The product $c_a^*c_b$ however, differs for these two eigenstates due to the $\pm$ signs before the square root in Eq.~(\ref{eq:c1c2}). We find
\be
{c_a^{(1)}}^*c^{(1)}_b = -\left[{c_a^{(2)}}^*c^{(2)}_b\right]^*, \label{eq:c_prod}
\ee
which indicates that $|g^{(1)}_{a,b}|\neq |g^{(2)}_{a,b}|$. We note that Eq.~(\ref{eq:c_prod}) is equivalent to
\begin{align}
(\theta^{(1)}-\pi/2)&=-(\theta^{(2)}-\pi/2),\quad\theta^{(1,2)}\in[0,\pi]\\
(\theta^{(1)}+\pi/2)&=-(\theta^{(2)}+\pi/2),\quad\theta^{(1,2)}\in(-\pi,0)
\end{align}
where $\theta^{(j)}$ is the relative phase between $c_a^{(j)}$ and $c_b^{(j)}$. In the $\cal PT$-broken phase $|g_0|<\kappa_0$, and we find
\be
|c_b|^2=\left(\frac{\kappa_0}{g_0}\pm\sqrt{\frac{\kappa_0^2}{g_0^2}-1}\right)^2|c_a|^2
\ee
from Eq.~(\ref{eq:c1c2}), which is clearly different from $|c_a|^2$.

For the nonlinear Hamiltonian given by Eq.~(\ref{eq:CMT_nonlinear}) in the main text, Eq.~(\ref{eq:c1c2}) still holds in the $\cal PT$-symmetric phase (where $g_b=g_a^*$) with a slight modification:
\be
c_b = \frac{\pm\sqrt{|g_a|^2-\kappa_0^2}-i\kappa_0}{g_a}c_a. \label{eq:c1c2_2}
\ee
It can be substituted back to the definition of $g_a(\varepsilon)$ given by Eq.~(\ref{eq:ga}) in the main text, which leads to a self-consistent equation
\be
g_a = g_0 + \varepsilon\beta\frac{\pm\sqrt{|g_a|^2-\kappa_0^2}-i\kappa_0}{2g_a} + \varepsilon\frac{\gamma}{2}. \label{eq:g_a}
\ee
$g_a$ (and $g_b$) can then be solved directly for a given nonlinearity strength $|\varepsilon|$, and the ``$\pm$" sign in front of the square root term in Eq.~(\ref{eq:g_a}) determines whether the system evolves along the path $H^{(1)}$ or $H^{(2)}$, i.e.,
\begin{align}
g_a^{(1)} &= g_0 + \varepsilon\beta\frac{\sqrt{|g_a^{(1)}|^2-\kappa_0^2}-i\kappa_0}{2g_a^{(1)}} + \varepsilon\frac{\gamma}{2}, \label{eq:g_a1}\\
g_a^{(2)} &= g_0 - \varepsilon\beta\frac{\sqrt{|g_a^{(2)}|^2-\kappa_0^2}+i\kappa_0}{2g_a^{(2)}} + \varepsilon\frac{\gamma}{2}. \label{eq:g_a2}
\end{align}

\subsection{Linear stability analysis}

\begin{figure}[t]
\includegraphics[clip,width=\linewidth]{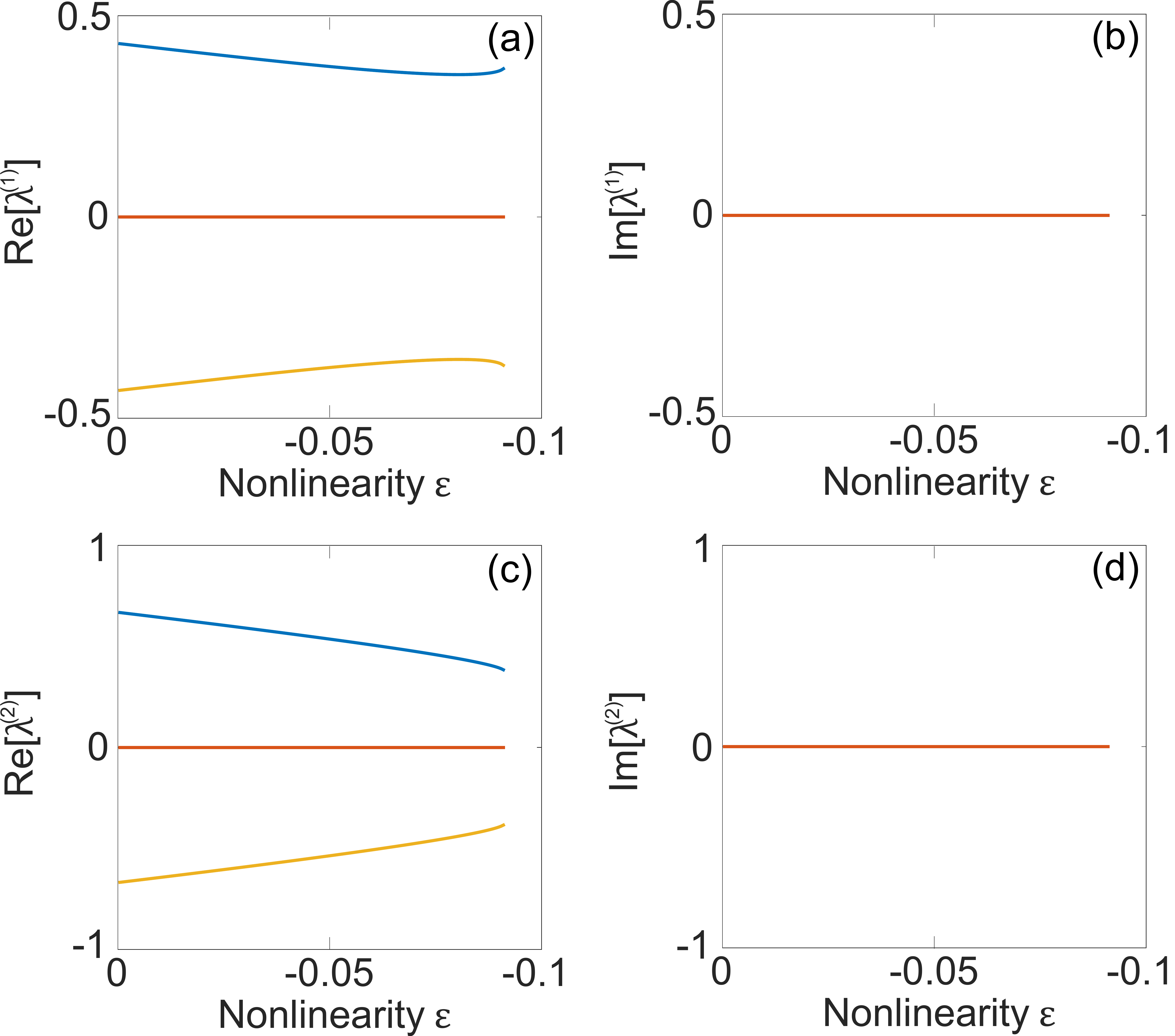}
\caption{Linear stability analysis of the nonlinear eigenstates $\psi^{(1,2)}(\varepsilon)$ in the $\cal PT$-symmetric phase of the example shown in Figs.~\ref{fig:nonlinear_cmt}(a) and (b). (a) and (b) Real and imaginary parts of the eigenvalues $\lambda$ of the stability matrix $M$ given by Eq.~(\ref{eq:stability}) for $\psi^{(1)}(\varepsilon)$. (c) and (d) Same as (a) and (b) but for $\psi^{(2)}(\varepsilon)$. In both cases there are four $\lambda$. Two of them are real with opposite nonzero values and the other two are zero.
}\label{fig:stability}
\end{figure}

The linear stability of the coupled mode theory can be analyzed by defining a perturbation $[\delta_a\;\delta_b]^T$ and studying the evolution of $[c_a+\delta_a\; c_b+\delta_b]^T$. We find
\begin{gather}
-i\frac{\partial}{\partial t}
\begin{pmatrix}
\delta_a \\
\delta_b
\end{pmatrix}
=(H+ \varepsilon M_1)
\begin{pmatrix}
\delta_a \\
\delta_b
\end{pmatrix}
+\varepsilon M_2
\begin{pmatrix}
\delta_a^* \\
\delta_b^*
\end{pmatrix}, \\
M_1=
\begin{pmatrix}
2|c_a|^2+\gamma c_a^*c_b & \beta c_a^*c_b \\
\beta c_b^*c_a & 2|c_b|^2+\gamma c_b^*c_a
\end{pmatrix}, \\
M_2=
\begin{pmatrix}
2c_a^2+\beta c_b^2+\gamma c_a c_b & 0\\
0 & 2c_b^2+\beta c_a^2+\gamma c_a c_b
\end{pmatrix},
\end{gather}
where $H$ is linearized about one of its eigenstates (e.g., $H^{(1,2)}$ in the $\cal PT$-symmetric phase). Since the eigenstate $[c_a\; c_b]^T$ is calculated for a given nonlinearity, we require the perturbation $[\delta_a\; \delta_b]^T$ to be orthogonal to $[c_a\; c_b]^T$ (otherwise it effectively changes the nonlinearity). This is a standard procedure \cite{Graefe,Castin} and can be done via the projection operator
\be
Q =
\begin{pmatrix}
1 - |c_a|^2 & -c_ac_b^* \\
- c_b c_a^*& 1 - |c_b|^2
\end{pmatrix},
\ee
and the stability of $[c_a\; c_b]^T$ is determined by the eigenvalues $\lambda$ of
\be
M =
\begin{pmatrix}
Q^{-1}(H + \varepsilon M_1)Q & \varepsilon Q^{-1} M_2 Q \\
-\varepsilon(Q^{-1} M_2Q)^* & -[Q^{-1}(H+\varepsilon M_1)Q]^*
\end{pmatrix}.\label{eq:stability}
\ee
As typical in such nonlinear systems, $\lambda$ are either real or form complex conjugate pairs, and the eigenstate of $H$ is stable if all $\lambda$ (four here) are real. We find the latter to be true for both $\psi^{(1,2)}$ in the $\cal PT$-symmetric phase of the example shown in Figs.~\ref{fig:nonlinear_cmt}(a) and (b) of the main text [see Fig.~\ref{fig:stability}].

\subsection{Location of the APT point}
\begin{figure}[t]
\includegraphics[clip,width=\linewidth]{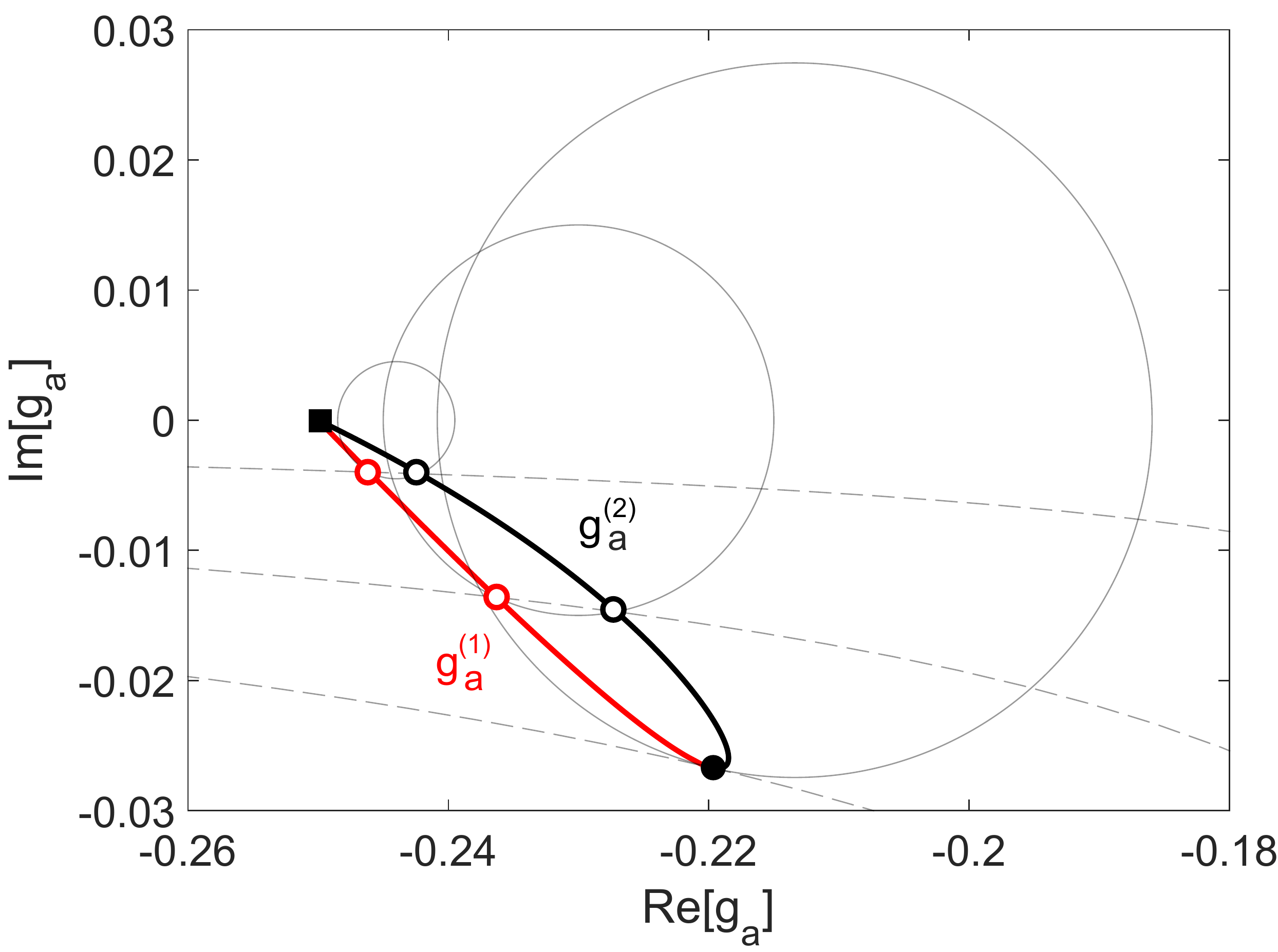}
\caption{(Color online) Trajectories of $g_a^{(1,2)}(\varepsilon)$ (thick solid lines) in the $\cal PT$-symmetric phase shown in Fig.~\ref{fig:nonlinear_cmt} of the main text. They are given by the intersections of the circle given by Eq.~(\ref{eq:g_a4}) (thin solid lines) and the hyperbola given by Eq.~(\ref{eq:g_a5}) (thin dashed lines) for a given $\varepsilon$. For the two small circles $\varepsilon=-0.015$ and $-0.05$. For $\varepsilon=-0.091$ the aforementioned circle and hyperbola become tangent to each other, and the tangent point gives the value of $g_a^{(1)}=g_a^{(2)}=-0.219 - 0.026i$ at the APT point (filled circle). The square shows $g_a^{(1)}(0)=g_a^{(2)}(0)=g_0=-0.025$.}\label{fig:g}
\end{figure}

To gain some analytical insights about the location of the APT point in terms of the nonlinearity $\varepsilon$, we rewrite Eq.~(\ref{eq:g_a}) as
\be
2\left(g_a - g_0 -\varepsilon\frac{\gamma}{2}\right)g_a= \varepsilon\beta\left(\pm\sqrt{|g_a|^2-\kappa_0^2}-i\kappa_0\right) . \label{eq:g_a3}
\ee
In the $\cal PT$-symmetric phase, this equation is equivalent to the following two:
\begin{align}
&\left(X - g_0 -\varepsilon\frac{\gamma}{2}\right)^2+Y^2 = \left(\frac{\varepsilon\beta}{2}\right)^2, \label{eq:g_a4}\\
&\left(2X - g_0 -\varepsilon\frac{\gamma}{2}\right)Y = -\frac{\varepsilon\beta\kappa_0}{2}, \label{eq:g_a5}
\end{align}
which are derived by taking the modulus and imaginary part of Eq.~(\ref{eq:g_a3}), respectively. We note that they hold for both $g_a^{(1,2)}$, where
$X$ and $Y$ represent the real and imaginary parts of $g_a^{(1,2)}$. $X$ and $Y$ are given by the intersections of a circle [Eq.~(\ref{eq:g_a4})] and a hyperbola [Eq.~(\ref{eq:g_a5})] in the complex plane for a given $\varepsilon$ (see Fig.~\ref{fig:g}). Note that these two curves become tangent to each other at a minimum $\varepsilon_\text{min}$ (with the maximum nonlinearity strength $|\varepsilon|_\text{max}$ since we consider $\varepsilon<0$ in the examples in the main text), below which these two curves do no intersect. In other words, $\varepsilon_\text{min}$ is where APT takes place, and it can be found by solving
\be
\left(G\hspace{-2pt}+\hspace{-2pt}\sqrt{G^2+2A^2}\right)\hspace{-4pt}\left(-3G\hspace{-2pt}+\hspace{-2pt}\sqrt{G^2+2A^2}\right)^3\hspace{-6pt} = 64A^2\kappa_0^2,
\ee
where $G\equiv g_0 + {\varepsilon_\text{min}\gamma}/{2}$ and $A\equiv2\varepsilon_\text{min}\beta$. This equation has multiple roots, and the one corresponding to the APT point in Fig.~\ref{fig:nonlinear_cmt} is $\varepsilon_\text{min}=-0.091$.

\subsection{Minimum of coupling in the $\cal PT$-symmetric phase}

In the main text we have mentioned that $\kappa_0$ is not just the cut-off of $|g_a(\varepsilon)|$ imposed by the $\cal PT$-symmetric phase; it is also the true minimum of $|g_a(\varepsilon)|$ as evidenced by the vanishing slope at the EP in Fig.~\ref{fig:nonlinear_cmt}(b). To prove this statement, one may attempt to find the minimum of $|g_a|$ in an optimization problem, with the two constraints Eqs.~(\ref{eq:g_a4}) and (\ref{eq:g_a5}) taken into account via Lagrange multipliers. However, its analytical form is very complicated, and here we provide a much simpler proof based on a perturbation analysis.

\begin{figure}[b]
\includegraphics[clip,width=\linewidth]{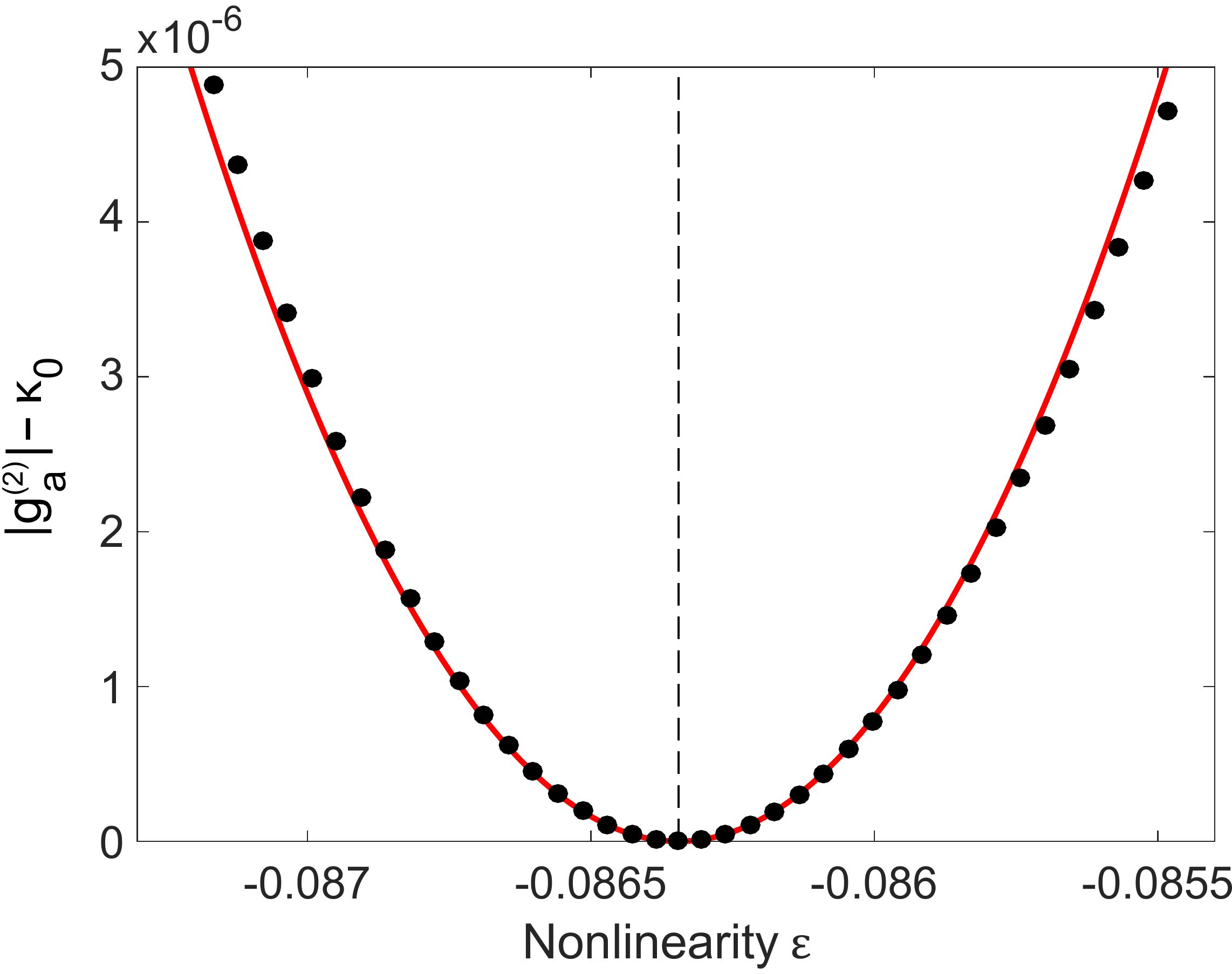}
\caption{(Color online) Behavior of the coupling $|g^{(2)}_a(\varepsilon)|$ near its EP in Fig.~\ref{fig:nonlinear_cmt}(b). Filled circles show the numerical solutions of Eq.~(\ref{eq:g_a}) and the solid line shows its approximation given by Eq.~(\ref{eq:g_min}). The value of the nonlinearity $\varepsilon$ at the EP is indicated by the dashed vertical line.}\label{fig:g_min}
\end{figure}

We first note that neither Eq.~(\ref{eq:g_a}) nor (\ref{eq:g_a3}) works in this approach due to the square root singularity. The latter, however, can be easily eliminated by rearranging Eq.~(\ref{eq:g_a3}) in the following way:
\be
\left[2\left(g_a - g_0 -\varepsilon\frac{\gamma}{2}\right)g_a+i\varepsilon\beta\kappa_0\right]^2= \varepsilon\beta(|g_a|^2-\kappa_0^2), \label{eq:g_a6}
\ee
where $|g_a|>\kappa_0$ in the $\cal PT$-symmetric phase. Next we define
\begin{align}
g_a \equiv g_\ep + g_1 \delta + g_2 \delta^2 + O(\delta^3) \quad
(\delta \equiv \varepsilon-\varepsilon_\ep), \label{eq:pert}
\end{align}
where $g_\ep$ (with $|g_\ep|=\kappa_0$) and $\varepsilon_\ep$ are the values of the coupling and nonlinearity at the EP. Because both sides of Eq.~(\ref{eq:g_a6}) vanish at the EP, the square on the left hand side then indicates that its leading order behavior is $O(\delta^2)$. The same needs to hold for the right hand side of Eq.~(\ref{eq:g_a6}), which is only possible if $g_1=0$:
\be
r.h.s = \beta^2\varepsilon_\ep^2(g_1g_\ep^*+g_1^*g_\ep)\delta + O(\delta^2).
\ee
Therefore, $g_a(\varepsilon)$, as well as its modulus, does not have a linear dependence on $\delta$ (and $\varepsilon$) near the EP. In other words, $\kappa_0$ is the true minimum of $|g_a(\varepsilon)|$ at $\varepsilon_\ep$ in the $\cal PT$-symmetric phase.

More quantitatively, the left and right hand sides of Eq.~(\ref{eq:pert}) are given by
\begin{align}
l.h.s &= (\gamma g_\ep - i\kappa_0\beta)^2 \delta^2 + O(\delta^3), \\
r.h.s &= \beta^2\varepsilon_\ep^2(g_2g_\ep^*+g_2^*g_\ep)\delta^2 + O(\delta^3),
\end{align}
and by equating their real parts we find $(g_2g_\ep^*+g_2^*g_\ep) =  \re{(\gamma g_\ep - i\kappa_0\beta)^2}/\beta^2\varepsilon_\ep^2$. Therefore,
\begin{align}
|g_a| &= \kappa_0 + \frac{1}{2\kappa_0}(g_2g_\ep^*+g_2^*g_\ep)\delta^2 + O(\delta^3) \nonumber \\
&= \kappa_0 + \frac{\re{(\gamma g_\ep - i\kappa_0\beta)^2}}{2\kappa_0\beta^2\varepsilon_\ep^2}\delta^2 + O(\delta^3),\label{eq:g_min}
\end{align}
which agrees nicely with the numerical solutions of Eq.~(\ref{eq:g_a}) (see Fig.~\ref{fig:g_min}).

\subsection{EP before the APT point}

\begin{figure}[t]
\includegraphics[clip,width=\linewidth]{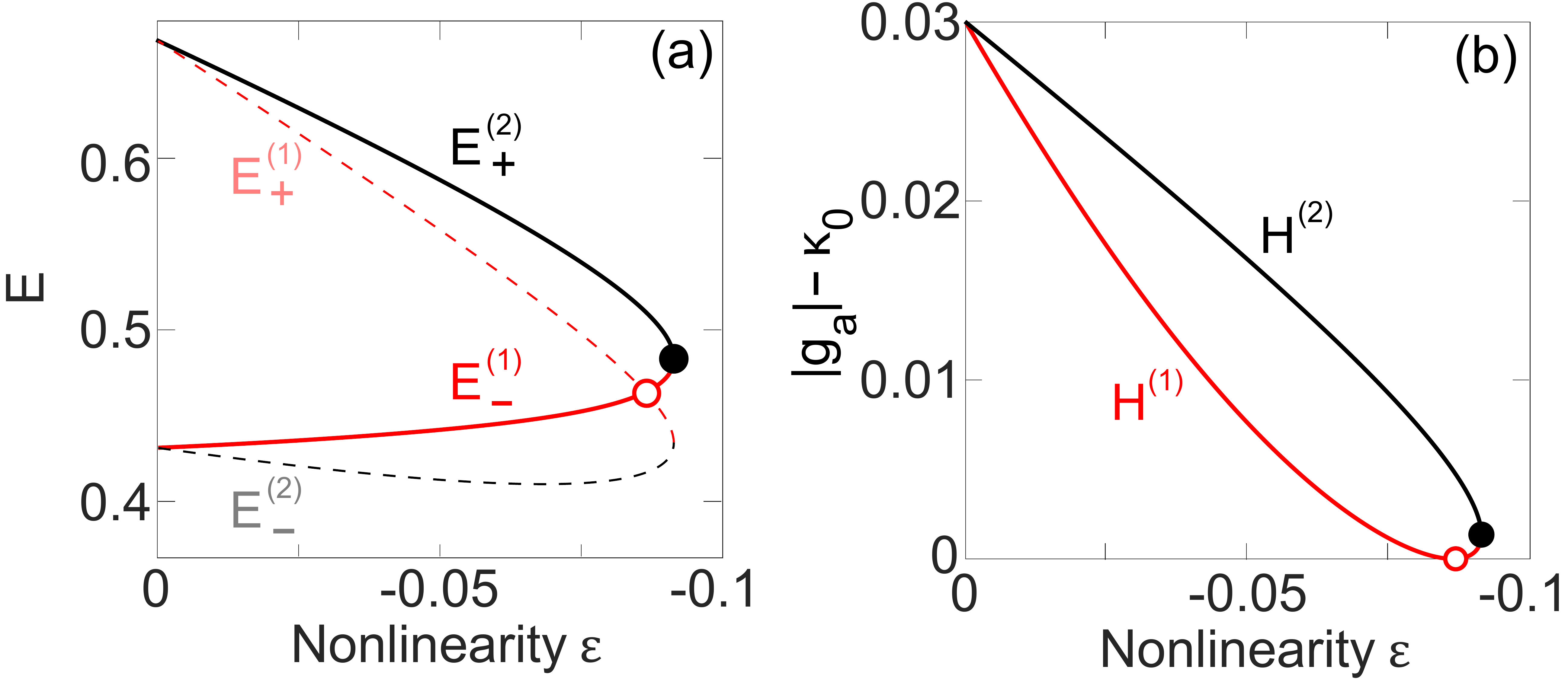}
\caption{(Color online) Another example of anomalous $\cal PT$ transition away from an EP. (a) and (b) are the same as those in Fig.~\ref{fig:nonlinear_cmt} except that the EP before the APT point is now on the nonlinear trajectory of $H^{(1)}$. The parameters are the same as in Fig.~\ref{fig:nonlinear_cmt} except for $\beta=-0.6$. }\label{fig:nonlinear_cmt2}
\end{figure}

In the example shown in Fig.~\ref{fig:nonlinear_cmt} of the main text, the EP before the APT point is along the trajectory of $H^{(2)}$. This EP can also appear along the trajectory of $H^{(1)}$, as we show in Fig.~\ref{fig:nonlinear_cmt2}. Which scenario appears depends on whether $|g^{(1)}_a|$ or $|g^{(2)}_a|$ is larger in the $\cal PT$-symmetric phase. For example, if $|g^{(1)}_a|<|g^{(2)}_a|$, then we find that $E^{(1)}_\pm$ are closer to each other than $E^{(2)}_\pm$ using Eq.~(\ref{eq:E_CMT_nonlinear}) in the main text, and the EP appears when $E^{(1)}_\pm$ becomes equal, i.e., on the trajectory of $H^{(1)}$. In fact, $E^{(j)}_\pm$ only depends on the absolute value of $g^{(j)}_a$ (and $g^{(j)}_b$). Therefore, by noting that
\be
g_a^{(2)}(-\beta) = \left[g_a^{(1)}(\beta)\right]^*
\ee
using Eqs.~(\ref{eq:g_a1}) and (\ref{eq:g_a2}),
we find that the values of $E^{(1)}_\pm$ are exchanged with $E^{(2)}_\pm$ when we flip the sign of $\beta$. In other words, the four curves in Fig.~\ref{fig:nonlinear_cmt2}(a) are identical with those in Fig.~\ref{fig:nonlinear_cmt}(a) but labeled differently. However, it is important to note that it is the lower lobe that corresponds to the two nonlinear eigenvalues of $H$ in Fig.~\ref{fig:nonlinear_cmt}, while it is the upper lobe that gives the two nonlinear eigenvalues of $H$ in Fig.~\ref{fig:nonlinear_cmt2}. In other words, the two sets of nonlinear eigenvalues, in systems with $\pm\beta$, crosses at the EP. We have used this property in the conclusion of the main text as a way to distinguish APT and a standard $\cal PT$ transition in a nonlinear system.

\subsection{Coupled-mode theory for a periodic system}

For a given wave number $k$, the modes of the Hermitian periodic potential $V_R(x)$ are given by the Bloch wave functions $\varphi_i(x;k)\exp(ikx)$, and $\varphi_i(x;k)$ are determined by
\be
\left[-\frac{\partial^2}{\partial x^2}  -2ik\-\frac{\partial}{\partial x} - k^2 + V_R(x)\right]\varphi_i(x;k) = E_i\varphi_i(x;k).\nonumber
\ee
We note that the corresponding energy eigenvalue $E_i$ is real.
It is straightforward to show that $\langle i|j\rangle\equiv\langle\varphi_i(x;k)|\varphi_j(x;k)\rangle=\delta_{ij}$ in the absence of degeneracy. $\langle \cdot | \cdot \rangle$ denotes the Hermitian inner product as usual.
In addition, the equation above is invariant upon the parity operation $x\rightarrow-x$ and taking the complex conjugate (note again that $E_i$ is real). Therefore, in principle we can find $\varphi_i(x;k)=\varphi_i^*(-x;k)$. Nevertheless, the global phase of $\varphi_i(x;k)$ is undetermined by its normalization $\langle i|i\rangle=1$. Thus we find
\be
\varphi_i(x;k)=\varphi_i^*(-x;k)\exp(2i\theta_i)
\ee
instead in general, where $\theta_i$ is the phase of $\varphi_i(x=0;k)$. %We then find $|\varphi_i(x;k)|^2=|\varphi_i^*(-x;k)|^2$, i.e., the intensity of all modes is symmetric about $x=0$.

We formulate a coupled-mode theory using modes $\varphi_{g,e}(x;k)$ in the first two bands of the Hermitian potential $V_R(x)$. As we shall see, a convenient choice is to set $\theta_g=0$ and $\theta_e=\pi/2$, leading to $\varphi_g(x;k)=\varphi_g^*(-x;k)$ and $\varphi_e(x;k)=-\varphi_e^*(-x;k)$. The presence of $V_I(x)$ (and $\xi|\psi|^2$) in principle couples modes of the same wave number $k$ in all bands, but the coupling is the strongest for modes in neighboring bands, and we find that the inclusion of $\varphi_{g,e}(x;k)$ is sufficient to demonstrate APT.

%With $\theta_g=0$, we have $\varphi_g(x;k)=\varphi_g^*(-x;k)$, or equivalently,
%\begin{gather}
%\re{\varphi_g(x)}=\re{\varphi_g(-x)},\\
%\im{\varphi_g(x)}=-\im{\varphi_g(-x)}.
%\end{gather}
%With $\theta_e=0$, we have $\varphi_e(x;k)=-\varphi_e^*(-x;k)$, or equivalently,
%\begin{gather}
%\re{\varphi_e(x)}=-\re{\varphi_e(-x)},\\
%\im{\varphi_e(x)}=\im{\varphi_e(-x)}.
%\end{gather}
%Therefore, $\langle \re{\varphi_g}|V_I|\im{\varphi_e}\rangle=\langle \im{\varphi_g}|V_I|\re{\varphi_e}\rangle=0$, because their integrands are odd functions about $x=0$. As a result, we find $\langle g|V_I|e\rangle=\langle\re{\varphi_g}|V_I|\re{\varphi_e}\rangle+\langle\im{\varphi_g}|V_I|\im{\varphi_e}\rangle$ is real, and so does $\langle e|V_I|g\rangle=\langle g|V_I|e\rangle$.

The basis of our coupled mode theory is chosen as
\be
\varphi_{a,b}(x;k) = \frac{1}{\sqrt{2}} (\varphi_g(x;k)\pm\varphi_e(x;k)),
\ee
which satisfy $\langle a|b \rangle=0$ and $\langle a|a \rangle=\langle b|b \rangle=1$.
With the phase conventions of $\varphi_{e,g}$ chosen above, the following relation also holds:
\be
\varphi_a(-x;k) %= (\varphi_g(-x;k) + \varphi_e(-x;k))/\sqrt{2}
= \frac{\varphi_g^*(x;k) - \varphi_e^*(x;k)}{\sqrt{2}}
=\varphi_b^*(x;k). \label{eq:v1v2}
\ee
We then find that $\langle a|V_I|a\rangle=-\langle b|V_I|b \rangle$:
\begin{align}
\langle a|V_I|a \rangle &= \int_{-L/2}^{L/2} \varphi_a^*(x;k)V_I(x;k)\varphi_a(x;k)\,dx\nonumber\\
&=\int_{-L/2}^{L/2} \varphi_a^*(-x;k)V_I(-x;k)\varphi_a(-x;k)\,dx\nonumber\\
&=\int_{-L/2}^{L/2} \varphi_b(x;k)[-V_I(x;k)]\varphi_b^*(x;k)\,dx\nonumber\\
&=-\langle b|V_I|b\rangle.\label{eq:kappa}
\end{align}
In the second step above we have performed a simple coordinate transformation ($x\rightarrow-x$), and in the third step we have used the relation (\ref{eq:v1v2}).
We define this expectation value as $\kappa_0$ to represent the gain and loss strength, which is real using the definition of the Hermitian inner product.

The effective Hamiltonian of the periodic system for the first two bands can then be written as
\be
H = \begin{pmatrix}
E_0 + i\kappa_0 & g \\
g & E_0 - i\kappa_0
\end{pmatrix}
+
\xi
\begin{pmatrix}
N_{a}   & L_{a}\\
L_{b} & N_{b}
\end{pmatrix}.\label{eq:CMT_periodic}
\ee
$E_0$ and the linear coupling $g$ are given by $(E_e\pm E_g)/2$, respectively.
We note that similar to Eq.~(\ref{eq:kappa}), we find $\langle a|V_I|b\rangle=-\langle a|V_I|b\rangle=0$, $\langle b|V_I|a\rangle=-\langle b|V_I|a\rangle=0$, which would have appeared in the off-diagonal elements of the linear part of $H$ in Eq.~(\ref{eq:CMT_periodic}). The nonlinear terms $N_a,L_a$ in this Hamiltonian are given by
\begin{align}
N_a&\equiv\langle aa|aa \rangle|c_a|^2+\langle ab|aa\rangle c_ac_b^* + 2\langle ab|ab\rangle|c_b|^2,\label{eq:N}\\
L_a&\equiv\langle ab|bb \rangle|c_b|^2+\langle aa|bb\rangle c_a^*c_b + 2\langle aa|ab\rangle|c_a|^2,\label{eq:L}
\end{align}
and $N_b,L_b$ are similarly defined with the subscripts $a$ and $b$ in these expressions exchanged. The quartic inner product here is defined by
\be
\langle ij|i'j' \rangle\equiv\hspace{-3pt}\int_{-L/2}^{L/2} \varphi_i^*(x;k)\varphi_j^*(x;k)\varphi_{i'}(x;k)\varphi_{j'}(x;k)\,dx, \nonumber
\ee
from which we see immediately that $\langle aa|aa\rangle$, $\langle bb|bb \rangle$, $\langle ab|ab \rangle$ are real by definition. In addition, we find $\langle aa|aa\rangle=\langle bb|bb \rangle$ using the relation (\ref{eq:v1v2}):
\begin{align}
\langle aa|aa \rangle &= \int_{-L/2}^{L/2} \varphi_a^*(x;k)\varphi_a^*(x;k)\varphi_a(x;k)\varphi_a(x;k)dx\nonumber\\
&=\int_{-L/2}^{L/2} \varphi_a^*(-x;k)\varphi_a^*(-x;k)\varphi_a(-x;k)\varphi_a(-x;k)dx\nonumber\\
&=\int_{-L/2}^{L/2} \varphi_b(x;k)\varphi_b(x;k)\varphi_b^*(x;k)\varphi_b^*(x;k)dx\nonumber\\
&=\langle bb|bb\rangle.\nonumber
\end{align}
Similarly, we find that $\langle ab|aa\rangle$ and $\langle ab|bb\rangle$ are complex conjugate of each other:
\begin{align}
\langle ab|aa\rangle &= \int_{-L/2}^{L/2} \varphi_a^*(x;k)\varphi_b^*(x;k)\varphi_a(x;k)\varphi_a(x;k)dx\nonumber\\
&=\int_{-L/2}^{L/2} \varphi_a^*(-x;k)\varphi_b^*(-x;k)\varphi_a(-x;k)\varphi_a(-x;k)dx\nonumber\\
&=\int_{-L/2}^{L/2} \varphi_b(x;k)\varphi_a(x;k)\varphi_b^*(x;k)\varphi_b^*(x;k)dx\nonumber\\
&=\left[\int_{-L/2}^{L/2} \varphi_b^*(x;k)\varphi_a^*(x;k)\varphi_b(x;k)\varphi_b(x;k)dx\right]^*\nonumber\\
&=\langle ab|bb\rangle^*.\nonumber
\end{align}
Therefore, we find that $N_a\equiv\langle aa|aa\rangle|c_a|^2+\langle ab|aa\rangle c_ac_b^* + 2\langle ab|ab\rangle|c_b|^2$ and $N_b\equiv\langle bb|bb\rangle|c_b|^2+\langle ab|bb\rangle c_bc_a^* + 2\langle ab|ab\rangle|c_a|^2$ are complex conjugate of each other when $|c_a|=|c_b|$. And finally, we note $L_a\equiv\langle ab|bb\rangle|c_b|^2+\langle aa|bb\rangle c_a^*c_b+2\langle aa|ab\rangle|c_a|^2$ and $L_b\equiv\langle ab|aa\rangle|c_a|^2+\langle bb|aa\rangle c_b^*c_a+2\langle bb|ab\rangle|c_b|^2$ are also complex conjugate of each other when $|c_a|=|c_b|$, once we realize that $\langle aa|bb\rangle=\langle bb|aa\rangle^*$ by the definition of the Hermitian inner product and $\langle aa|ab\rangle=\langle ab|aa\rangle^*=(\langle ab|bb\rangle^*)^*=\langle bb|ab\rangle^*$.

\begin{figure}[t]
\includegraphics[clip,width=\linewidth]{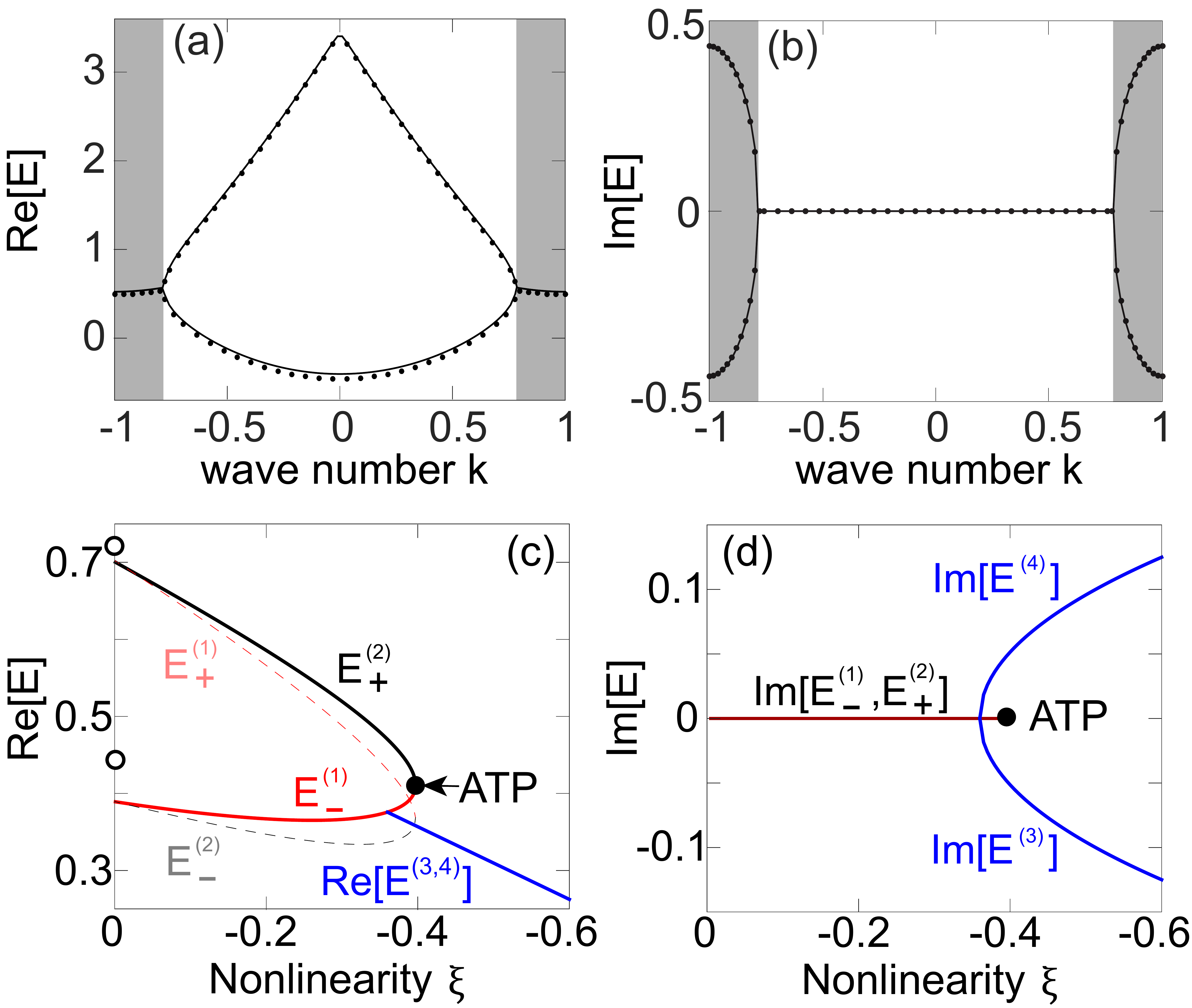}
\caption{(a,b) Same as Figs.~\ref{fig:periodic}(a) and \ref{fig:periodic}(b) but showing the whole Brillouin zone. The dots show the linear bands calculated using the coupled mode theory (\ref{eq:CMT_periodic}) with $\xi=0$. (c,d) Same as Figs.~\ref{fig:periodic}(c) and \ref{fig:periodic}(d) but reproduced using the coupled-mode theory (\ref{eq:CMT_periodic}) with $\xi\leq0$. Open circles in (c) show the energies of the two bands at $k=0.77$ by solving Eq.~(\ref{eq:paraxial}) in the main text directly.
}\label{fig:periodic_cmt}
\end{figure}

In conclusion, we find that the nonlinear effective Hamiltonian given by Eq.~(\ref{eq:CMT_periodic}) is $\cal PT$-symmetric when $|c_1|=|c_2|$, i.e., ${\cal PT}H{\cal PT}=H$. In addition, it depends on the relative phase of $c_a$ and $c_b$, which leads to two distinct trajectories of $H$ that depend on the nonlinear modal index $j$, just like the simpler form of $H$ given by Eq.~(\ref{eq:CMT_nonlinear}) in the main text.

This coupled mode theory agrees well with the direct numerical solutions of the paraxial equation (\ref{eq:paraxial}) in the linear case, as we show in Figs.~\ref{fig:periodic_cmt}(a) and \ref{fig:periodic_cmt}(b). For the pair of modes at $k=0.77$ shown in Fig.~\ref{fig:periodic}(c) and \ref{fig:periodic}(d), we find $E_g=0.0208$, $E_e=1.0692$, $\kappa_0=0.5006$, $\langle aa|aa\rangle = \langle bb|bb\rangle = 0.4791$, $\langle ab|aa\rangle=\langle bb|ab\rangle=0.0048 - 0.0088i=\langle ab|bb\rangle^*=\langle aa|ab\rangle^*$, $\langle ab|ab\rangle=0.1588$, and $\langle aa|bb\rangle=-0.0848 + 0.1343i=\langle bb|aa\rangle^*$. This coupled mode theory reproduces qualitatively the APT shown in Fig.~\ref{fig:periodic} in the main text [see Fig.~\ref{fig:periodic_cmt}(c)], and we note that a deviation occurs due to the neglect of the coupling to higher order bands: the EP now appears along the path of $H^{(1)}$ instead of $H^{(2)}$. The small deviation in the coupled theory can be seen in the linear case as well, as we show in Fig.~\ref{fig:periodic_cmt}(c) at $\xi=0$.

\end{document}